\documentclass[seceq]{ptptex}

\usepackage{graphicx}
\usepackage{amsmath}

\title{
Distribution of Resonant Eigenvalues of Quantum Potential Scattering%
}

\author{
Naomichi \textsc{Hatano}\footnote{E-mail: hatano@iis.u-tokyo.ac.jp}
}

\inst{
Institute of Industrial Science, University of Tokyo,\\
4-6-1 Komaba, Meguro, Tokyo 153-8505, Japan\\
}

\abst{
We formulate the Born approximation for finding resonance poles in the complex plane for potential scattering problems.
Using the method, we study the distribution of resonance poles for several scattering potentials.
In particular, we find for an exponential potential with a cutoff that the cutoff generates an infinite series of extra resonance poles below and along the real axis, which would not exist without the cutoff.
We also find for a Gaussian potential that the series of resonance poles approach the imaginary axis of the complex energy plane from left.
In other words, the real parts of the resonant eigenenergis are all negative.
}

\begin{document}

\maketitle

\section{Introduction}
\label{sec1}

Resonance is a phenomenon prevalent all over physics~\cite{Newton82,Kukulin89,Brandas90}.
Quantum resonance in particular was studied intensively perhaps first in the field of nuclear physics~\cite{Bethe35,Bethe36,Hulthen42a,Hulthen42b,Jost47,Wigner55,Nakanishi58,Humblet61,Rosenfeld61,Humblet62,Humblet64a,Jeukenne64,Humblet64b,Mahaux65,Rosenfeld65,Bohm89,Homma97,Myo97,Myo98,Masui99,Amrein87,Carvalho02,Ahmed04,Kelkar04,Jain05,Amrein06,Rotter09} and maybe later in atomic and molecular physics~\cite{Smith62,Burke62,Schulz64,Ho77,Agostini79,Freeman87,Moiseyev90,Peskin92,Peskin93,Moiseyev97,Brisker08,Moiseyev08}.
In recent years, quantum resonance is attracting much attention in condensed-matter physics as well~\cite{Tekman93,Gores00,Zacharia01,Clerk01,Racec01,Kang01,Torio02,Kobayashi02,Kobayashi03,Kobayashi04,Sato05,Kim03,Babic04,Lu05,Chakrabarti06,Franco06,Joe07,Fujimoto08}, thanks to the development of nanotechnology.

Many studies on quantum resonance focus on resonant states with long lifetimes, or in other words, resonance poles close to the real axis in the complex energy plane.
Resonant states far from the real axis tend to be ignored on the ground that they are very unstable and hardly observable in experiments.

A recent study~\cite{Sasada05,Sasada09}, however, revealed that resonance poles far from the real axis do affect resonance peaks on the real axis in the form of the Fano asymmetry.
A conventional (and perhaps somewhat phenomenological) understanding of the Fano asymmetry is that we have an asymmetric resonance peak when a bound state and an energy continuum are coupled~\cite{Fano61}.
In contrast, the above-mentioned study~\cite{Sasada05,Sasada09} microscopically expanded the transmission amplitude with respect to resonant states and showed that quantum interference among resonant states and bound states results in several types of asymmetry in resonance peaks.
In particular, when we have two neighboring resonance poles, one close to the real axis (with a long lifetime and a narrow peak) and one far from the real axis (with a short lifetime and a broad peak), interference between these poles make the peak of the former resonance strongly asymmetric.
This opens up the possibility of experimentally observing resonance poles far from the real axis.

We thereby emphasize in the present paper the importance of studying the distribution of the resonance poles over the entire complex plane.
There have been quite a few studies that obtained the distribution of the resonance poles mainly for exactly solvable problems of potential scattering~\cite{Eckart30,Aly65,Jost51,Vogt54,Corinaldesi56,Nussenzveig59,Fivel60,Bhattacharjie62,Wojtczak63,Bose64,Spector64,Fuda71,Bahethi71,Bawin74,Doolen78,Rittby82,Ginocchio84,Alhassid85,Colbert86,Benjamin86}.
In contrast, we will here develop a Born approximation for finding resonance poles in the entire complex plane.
The approximation is applicable to a wide class of solvable and unsolvable scattering potentials.

Let us briefly summarize the results of the Born approximation here.
We will show the following in the present paper:
\begin{enumerate}
\renewcommand{\labelenumi}{(\roman{enumi})}
\item For exponential potentials, all resonance poles are on the negative imaginary axis of the complex wave-number plane, or on the negative real axis of the second Riemann sheet of the complex energy plane.
\item For exponential potentials with oscillatory modulations, the resonance poles which would be on the negative imaginary axis of the complex wave-number plane without oscillation, now split into two and shift sideways into the third and fourth quadrants of the complex wave-number plane.
\item For exponential potentials with a discontinuous cutoff, an infinite series of extra resonance poles appear below and along the real axis of the complex wave-number plane.
It is therefore dangerous to introduce a cutoff carelessly perhaps for numerical calculations.
\item For Gaussian potentials, a series of the resonance poles approach to the line of argument $-\pi/4$ in the complex wave-number plane, or to the negative imaginary axis in the complex energy plane.
\end{enumerate}
We believe that the third and fourth results are of particular importance.
We will then in the final section come up with two conjectures on the distribution of the resonance poles for a wider class of scattering potentials.

For readers' benefit, we in Sec.~\ref{sec15} summarize the basic knowledge on the resonant state.
We then introduce the Born approximation for resonant states in Sec.~\ref{sec2}.
Each of Secs.~\ref{sec3}--\ref{sec6} presents the afore-mentioned results (i)--(iv) in this order, respectively.
In the final section, we summarize the paper and present two conjectures on the distribution of the resonance poles.
We add two appendices, one to summarize the formulation of the Jost function in one dimension, the other to review the Jost-function method, a numerical method of finding resonance poles.

\section{Definition of resonant states}
\label{sec15}

Before going into the points of the present paper, we here in the present section give a brief review of the definition of resonant states.
Resonant states can be described statically or dynamically.
A dynamical view of a resonant state may be as follows;
a wave packet with a distribution of energy comes into a potential and gets captured in a quasi-bound state for a long time before getting scattered away eventually.
We do not take this dynamical view in the present paper at all.

A static view of a resonant state taken in many textbooks may be a pole of the $S$-matrix in the complex energy plane.
In fact, we can equivalently define a resonant state as an eigenstate of the stationary Schr\"{o}dinger equation
\begin{align}\label{eq1-5}
-\frac{d^2}{dx^2}\phi(k;x)+V(x)\phi(k;x)=k^2\phi(k;x),
\end{align}
where we put $\hbar^2/2m$ to unity.
The eigenenergy $E$ of the state is given by the eigen-wave-number $k$ as $E=k^2$.

The definition of a resonant state as an eigenstate is achieved under the boundary condition that only out-going waves exist~\cite{Siegert39,Landau77,Hatano08},
which is often called the Siegert condition.
Suppose that we have a scattering eigenfunction in the form
\begin{align}\label{eq1-10}
\phi(k;x)=
\begin{cases}
A(k)e^{ikx}+B(k)e^{-ikx} & \quad\mbox{for $x\to-\infty$}, \\
C(k)e^{ikx} & \quad\mbox{for $x\to+\infty$}.
\end{cases}
\end{align}
Matching conditions at the scattering potential yield in the coefficients $A$, $B$ and $C$ the dependence on the wave number $k$ (and hence on the energy $E=k^2$).
In order to have out-going waves only, we seek zeros of the coefficient $A(k)$ of the in-coming wave,
\begin{align}\label{eq1-20}
A(k_\textrm{res})=0,
\end{align}
which is indeed equivalent to seeking poles of the $S$-matrix, or the transmission amplitude $C(k)/A(k)$.

We summarize here some properties of thus-defined resonant states (Table~\ref{tab1});
see Ref.~\citen{Hatano08} for details.
\begin{table}[t]
\caption{Summary of definitions of resonant states, anti-resonant states, bound states, and anti-bound states.
The eigenenergy $E_\textrm{res}$ is given by the eigen-wave-number $k_\textrm{res}$ as $E_\textrm{res}={k_\textrm{res}}^2$.}
\label{tab1}
\begin{center}
\begin{tabular}{lccccc}
\hline
Solutions of Eq.~(\ref{eq1-20}) 
& $\mathop{\textrm{Re}} k_\textrm{res}$ & $\mathop{\textrm{Im}} k_\textrm{res}$
& $\mathop{\textrm{Re}} E_\textrm{res}$ & $\mathop{\textrm{Im}} E_\textrm{res}$
& Riemann sheets \\
\hline\hline
Resonant state & positive & negative & any & negative & second \\
\hline
Anti-resonant state & negative & negative & any & positive & second \\
\hline
Bound state & zero & positive & negative & zero & first \\
\hline
Anti-bound state & zero & negative & negative & zero & second \\
\hline
\end{tabular}
\end{center}
\end{table}
Solutions of Eq.~(\ref{eq1-20}) in the fourth quadrant of the complex $k$ plane are indeed called ``resonant states," whereas solutions in the third quadrant are called ``anti-resonant states."
(In some context, ``anti-resonance" means a dip rather than a peak in the energy-dependence of the transmission probability.
It is not related to the anti-resonant state defined here.)
In a resonant state, particles trapped in the scattering potential leak to the infinity.
An anti-resonant state is a time-reversal state of a resonant state.
They always appear in pair.
In terms of the energy, a resonant state is in the lower half of the second Riemann sheet of the complex $E$ plane, whereas an anti-resonant state is in the upper half of the same sheet.
Therefore, the particle number in a resonant state decays exponentially in time, while the particle number in an anti-resonant state grows exponentially.

Solutions of Eq.~(\ref{eq1-20}) in the upper half of the $k$ plane must exist on the imaginary axis.
They are indeed bound states.
To see this, go back to Eq.~(\ref{eq1-10}).
For a solution of Eq.~(\ref{eq1-20}), the wave function~(\ref{eq1-10}) reduces to
\begin{align}\label{eq1-100}
\phi(k_\textrm{res};x)=e^{ik_\textrm{res}|x|},
\end{align}
which further reduces to 
\begin{align}\label{eq1-110}
\phi(k_\textrm{res}=i\beta;x)=e^{-\beta|x|}
\end{align}
for a pure imaginary solution $k=i\beta$.
This is indeed a bound state if $\beta>0$.
A bound state is on the negative part of the real axis of the first Riemann sheet of the complex $E$ plane.

Readers may then notice that the wave function~(\ref{eq1-100}) diverges as $|x|\to\infty$ for the solutions in the lower half of the $k$ plane, namely the resonant and anti-resonant states.
This may appear to be unphysical, but in fact indicates that the particles trapped in the scattering potential will go away to infinity eventually.
Indeed, we can show~\cite{Hatano08} that the divergence is essential for the particle-number conservation, by noting that the divergence is balanced with the exponential decay in time, $\exp(-t|\mathop{\textrm{Im}}E_\textrm{res}|)$.

For some potentials, a resonant state in the fourth quadrant and an anti-resonant state in the third quadrant collide and turn into two solutions on the negative part of the imaginary axis of the complex $k$ plane.
Such solutions are often called ``anti-bound states."~\cite{Ohanian74}
In terms of the energy, an anti-bound state is on the negative part of the real axis of the second Riemann sheet of the complex $E$ plane.

\section{Born approximation for resonant states}
\label{sec2}

We are now in position to formulate the Born approximation for finding resonant, anti-resonant and anti-bound states.
We here consider quantum-mechanical scattering due to symmetric potentials in the infinite one-dimensional space.
All eigenfunctions are hence classified into even and odd functions.
For spherical potentials in the three-dimensional space, we choose only odd eigenfunctions because the radial wave function must vanish at the origin of the semi-infinite radial coordinate.

We start with the Schr\"{o}dinger equation~(\ref{eq1-5}).
As described above, we assume $V(-x)=V(x)$.
In the corresponding Lippmann-Schwinger equation
\begin{align}\label{eq2-20}
\phi(k;x)=\phi_0(k;x)+\int_{-\infty}^\infty G(k;x-x')V(x')\phi(k;x')dx',
\end{align}
the retarded Green's function in one dimension is given by
\begin{align}\label{eq2-30}
G(k;x-x')=\frac{1}{2ik}e^{ik|x-x'|}
\end{align}
and the unperturbed wave function $\phi_0(k;x)$ is given by
\begin{align}\label{2-40}
\phi_0(k;x)=e^{ikx}+\sigma e^{-ikx},
\end{align}
where
\begin{align}\label{eq2-50}
\sigma=
\begin{cases}
+1 & \mbox{for even solutions,} \\
-1 & \mbox{for odd solutions.}
\end{cases}
\end{align}

In the Born approximation, we replace the perturbed wave function $\phi(k;x)$ in the integrand of Eq.~(\ref{eq2-20}) with the unperturbed wave function $\phi_0(k;x)$.
In the following, we will be interested only in the behavior in the limit $x\to\infty$.
We hence use only the part $x>x'$ in the retarded Green's function~(\ref{eq2-30}).
We thus use the formula
\begin{align}\label{eq2-60}
\lim_{x\to+\infty}\phi(k;x)
&=e^{ikx}+\sigma e^{-ikx}
+\frac{1}{2ik}\int_{-\infty}^\infty e^{ik(x-x')} V(x')\left(e^{ikx'}+\sigma e^{-ikx'}\right)dx'
\nonumber\\
&=e^{ikx}+\sigma e^{-ikx}
+\frac{e^{ikx}}{2ik}\int_{-\infty}^\infty V(x')\left(1+\sigma e^{-2ikx'}\right)dx'.
\end{align}

After obtaining the asymptotic form~(\ref{eq2-60}) for a specific potential $V(x)$, we will seek resonance poles by solving Eq.~(\ref{eq1-20}).
More specifically, since we classify eigenfunctions into even and odd functions, we symmetrize and antisymmetrize the scattering state in the form
\begin{align}\label{eq2-70}
\phi(k;x)=
\begin{cases}
\sigma A(k)e^{ikx}+\left(\sigma B(k)+C(k)\right)e^{-ikx} & \mbox{for $x\to-\infty$,} \\
A(k)e^{-ikx}+\left(B(k)+\sigma C(k)\right)e^{ikx} & \mbox{for $x\to+\infty$,}
\end{cases}
\end{align}
or after changing the normalization
\begin{align}\label{eq2-80}
\phi(k;x)=
\begin{cases}
\displaystyle
\sigma e^{-ikx}+\sigma\frac{A(k)}{B(k)+\sigma C(k)}e^{ikx} & \mbox{for $x\to-\infty$,} \\
\displaystyle
e^{ikx}+\frac{A(k)}{B(k)+\sigma C(k)}e^{-ikx} & \mbox{for $x\to+\infty$.}
\end{cases}
\end{align}
Therefore, the Siegert condition that we have out-going waves only gives the resonance equation in the form
\begin{align}\label{eq2-90}
\frac{A(k_\textrm{res})}{B(k_\textrm{res})+\sigma C(k_\textrm{res})}=0.
\end{align}
This produces the resonance poles $k=k_\textrm{res}$ in the complex wave-number plane and hence $E=E_\textrm{res}\equiv {k_\textrm{res}}^2$ in the complex energy plane.

\section{Exponential potential}
\label{sec3}

In the present section, we discuss the resonance spectrum for a potential with one or multiple exponential functions.
We start with one exponential,
\begin{align}\label{eq3-10}
V(x)=-V_0 e^{-\kappa_0 |x|},
\end{align}
where $V_0>0$ and $\kappa_0>0$;
within the Born approximation, the spectrum for the case of multiple exponential functions is simply a superposition of the case of one exponential function.

The single exponential~(\ref{eq3-10}) is an exactly solvable potential.
The Jost solution, which is defined by the behavior $e^{ik|x|}$ in the limit $|x|\to\infty$, is given by~\cite{Newton82,Bethe36,Jost47}
\begin{align}\label{eq3-1010}
f_+(k;x)=\exp\left[i(k/\kappa_0)\ln\left(V_0/{\kappa_0}^2\right)\right]
\Gamma\left(1-2iak/\kappa_0\right)
J_{-2ik/\kappa_0}\left(2\sqrt{V_0e^{-\kappa_0 x}/{\kappa_0}^2}\right),
\end{align}
where $J_\nu(z)$ is the Bessel function of the first kind.
The Jost function $F_+(k)$ (which should not be confused with the Jost solution) is defined as Eq.~(\ref{eqa-50}).
The resonance equation~(\ref{eq2-90}) in this case is expressed as Eq.~(\ref{eqa-150}).
In the present case in particular, the resonant states come from the following part of the equation:
\begin{align}\label{eq3-1030}
0=\begin{cases}
J'_{-2ik_\textrm{res}/\kappa_0}\left(2\sqrt{V_0}/\kappa_0\right) & \mbox{for even solutions,} \\
J_{-2ik_\textrm{res}/\kappa_0}\left(2\sqrt{V_0}/\kappa_0\right) & \mbox{for odd solutions.}
\end{cases}
\end{align}
These equations give solutions only on the imaginary axis, namely the bound states and the anti-bound states.
There are infinitely many anti-bound states with
\begin{align}\label{eq3-1040}
k_\textrm{res} \begin{cases}
\gtrsim -\frac{i}{2}n\kappa_0 & \mbox{for even solutions,} \\
\lesssim -\frac{i}{2}n\kappa_0 & \mbox{for odd solutions,}
\end{cases}
\end{align}
for large $|k_\textrm{res}|$.

We are going to show below that the Born approximation for the single exponential function~(\ref{eq3-10}) yields a pair of anti-bound states.
Although the Born approximation does not reproduce the fact that there are infinitely many anti-bound states, it does reproduce the fact that there are only anti-bound states in the lower half plane, no pairs of resonant and anti-resonant states.
This may be remarkable considering that the approximation is quite naive and furthermore the imaginary axis is where it is at its worst;
the Born approximation is considered to be accurate away from the imaginary axis, that is, when the particle is running fast.

Let us now start the Born approximation.
The formula~(\ref{eq2-60}) is followed by straightforward calculation for the potential~(\ref{eq3-10}), resulting
\begin{align}\label{eq3-20}
\lim_{x\to+\infty}
\phi(k;x)=\sigma e^{-ikx}
+e^{ikx}\left[1-\frac{V_0}{2ik}
\left(\frac{2}{\kappa_0}+\frac{\sigma}{\kappa_0-2ik}+\frac{\sigma}{\kappa_0+2ik}\right)
\right].
\end{align}
We can change the normalization of the wave function~(\ref{eq3-20}) as in Eq.~(\ref{eq2-80}), but we can do it in a more symmetric way.
Considering that singularities in the lower half of the complex $k$ plane give resonant states, we split the term proportional to $V_0$ in Eq.~(\ref{eq3-20}) into two and renormalize the wave function as
\begin{align}\label{eq3-30}
\lefteqn{\lim_{x\to+\infty}
\phi(k;x)}
\nonumber\\
&\simeq\sigma e^{-ikx}
+e^{ikx}\left[1-\frac{V_0}{2ik}
\left(\frac{1}{\kappa_0}+\frac{\sigma}{\kappa_0-2ik}\right)
\right]
\left[1-\frac{V_0}{2ik}
\left(\frac{1}{\kappa_0}+\frac{\sigma}{\kappa_0+2ik}\right)
\right]
\nonumber\\
&\propto\sigma e^{-ikx}
\left[1-\frac{V_0}{2ik}
\left(\frac{1}{\kappa_0}+\frac{\sigma}{\kappa_0-2ik}\right)
\right]^{-1}
+e^{ikx}\left[1-\frac{V_0}{2ik}
\left(\frac{1}{\kappa_0}+\frac{\sigma}{\kappa_0+2ik}\right)
\right]
\nonumber\\
&\simeq
\sigma e^{-ikx}
\left[1+\frac{V_0}{2ik}
\left(\frac{1}{\kappa_0}+\frac{\sigma}{\kappa_0-2ik}\right)
\right]
+e^{ikx}\left[1-\frac{V_0}{2ik}
\left(\frac{1}{\kappa_0}+\frac{\sigma}{\kappa_0+2ik}\right)
\right].
\end{align}
We thereby have a form that is symmetric with respect to $k\to -k$ (except for the factor $\sigma$).

The resonance equation~(\ref{eq2-90}) now reduces to
\begin{align}\label{eq3-40}
1+\frac{V_0}{2ik_\textrm{res}}
\left(\frac{1}{\kappa_0}+\frac{\sigma}{\kappa_0-2ik_\textrm{res}}\right)=0.
\end{align}
It is easy to see that, within the Born approximation, the resonance equation for a potential with multiple exponential functions
\begin{align}\label{eq3-50}
V(x)=-\sum_j V_j e^{-\kappa_j|x|}
\end{align}
is a simple extension of Eq.~(\ref{eq3-40}):
\begin{align}\label{eq3-60}
1+\sum_j\frac{V_j}{2ik_\textrm{res}}
\left(\frac{1}{\kappa_j}+\frac{\sigma}{\kappa_j-2ik_\textrm{res}}\right)=0.
\end{align}

It may be noteworthy that the same equation can be produced by the method described in Jost's paper~\cite{Jost47}.
For the potential of the form~(\ref{eq3-50}), Jost expanded the Jost solution as
\begin{align}\label{eq3-70}
f_+(k;x)=e^{ikx}\left(1+\sum_j\sum_{n=1}^\infty C_{jn}e^{-n\kappa_jx}\right)
\end{align}
for $x>0$ and obtained recursion relations for the coefficients $C_{jn}$.
It turns out that $C_{jn}=O({V_0}^n)$.
Therefore, the Born approximation is equivalent to neglecting the terms $n\geq2$.
As is mentioned above, the resonant states are given by the zeros of the Jost function defined by Eq.~(\ref{eqa-50}).
We can show after some algebra that Eq.~(\ref{eqa-150}) (with the approximation of neglecting the terms $n\geq2$) is indeed equivalent to Eq.~(\ref{eq3-60}).

\begin{figure}[t]
\centering
\begin{minipage}[t]{0.4\textwidth}
\includegraphics[width=\textwidth]{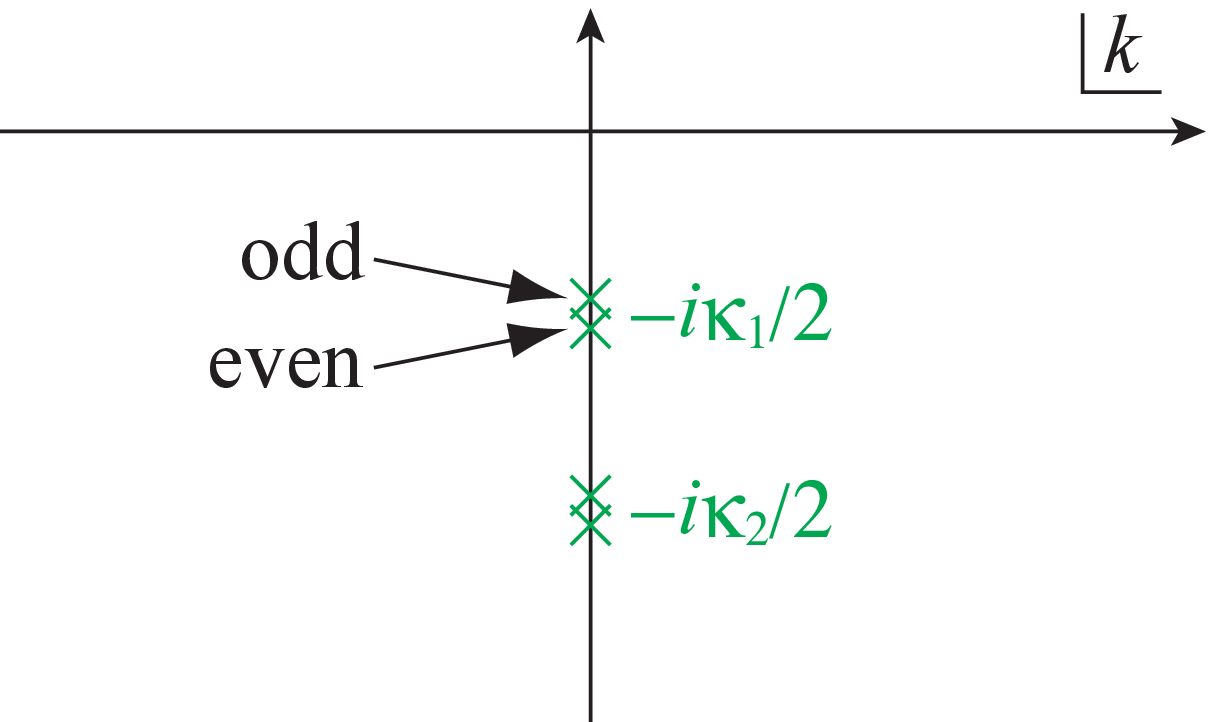}
\centering (a)
\end{minipage}
\hspace{0.08\textwidth}
\begin{minipage}[t]{0.4\textwidth}
\includegraphics[width=\textwidth]{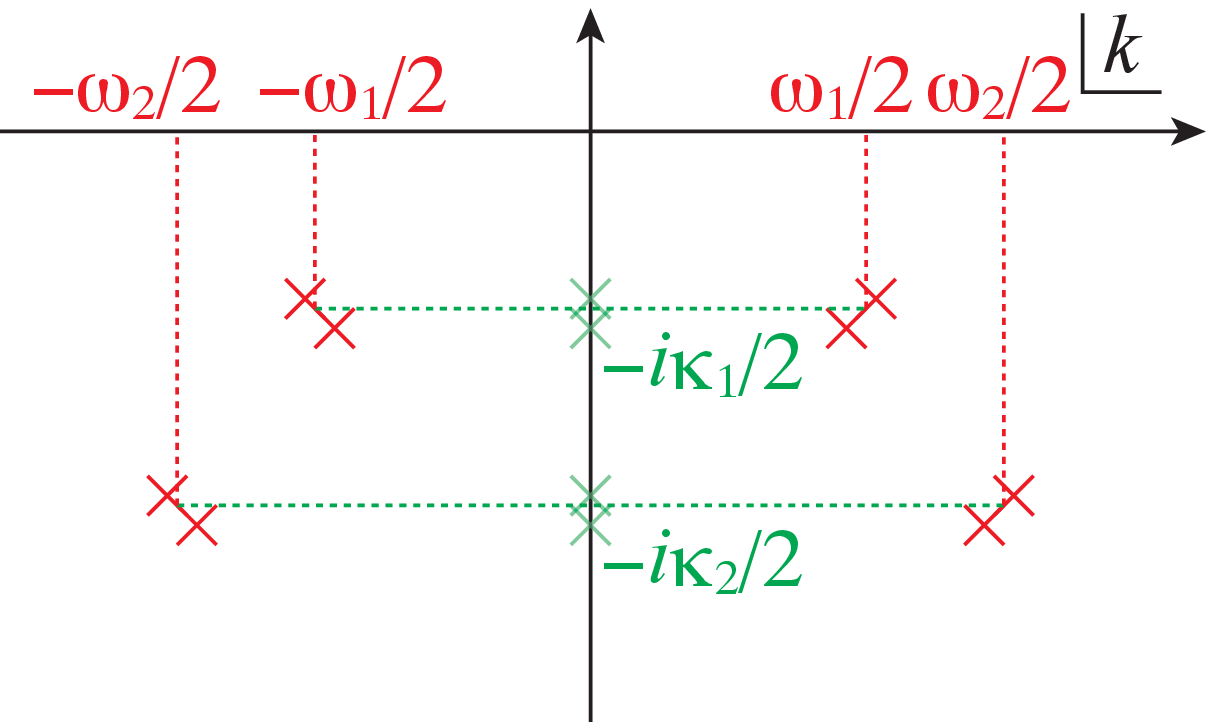}
\centering (b)
\end{minipage}
\caption{A schematic illustration of the resonance spectrum within the Born approximation: (a) for a potential with multiple exponentials, we have a pair of even and odd solutions corresponding to each exponential; (b) for a potential with multiple oscillatory exponentials, each solution splits into a resonant state and an anti-resonant state.}
\label{fig1}
\end{figure}

Now we solve the resonance equation~(\ref{eq3-40}).
We could solve it separately for $\sigma=+1$, when the equation is quadratic, and for $\sigma=-1$, when the equation is linear.
We here solve it more systematically.
We note that the dominant contribution of Eq.~(\ref{eq3-40}) comes from the term with the denominator $\kappa_0-2ik_\textrm{res}$.
We therefore have
\begin{align}\label{eq3-100}
2ik_\textrm{res}(2ik_\textrm{res}-\kappa_0)=\sigma V_0,
\end{align}
whose zeroth-order solution is $2ik_\textrm{res}\simeq \kappa_0$.
After putting $2ik_\textrm{res}=\kappa_0+\delta$, we then have
\begin{align}\label{eq3-110}
(\kappa_0+\delta)\delta=\sigma V_0.
\end{align}
Since $\delta=O(V_0)$, Eq.~(\ref{eq3-110}) actually reduces to $\kappa_0\delta=\sigma V_0$ in the first order of $V_0$.
We thereby arrive at a pair of solutions on the negative imaginary axis:
\begin{align}\label{eq3-120}
k_\textrm{res}=-\frac{i}{2}\left(\kappa_0+\sigma\frac{V_0}{\kappa_0}\right).
\end{align}
For the potential with multiple exponential functions, Eq.~(\ref{eq3-50}), there is a pair of solutions for each exponential:
\begin{align}\label{eq3-130}
k_\textrm{res}=-\frac{i}{2}\left(\kappa_j+\sigma\frac{V_j}{\kappa_j}\right)
\qquad\mbox{for each $j$}.
\end{align}
This is illustrated in Fig.~\ref{fig1}(a);
we have a pair of even and odd solutions on the imaginary axis for each exponential in the potential~(\ref{eq3-50}).
Note that there are no exact solutions for potentials with multiple exponential functions.

We have thus showed within the Born approximation that the resonant solutions are only on the negative imaginary axis for potentials with one or more exponentials.
At the end of the present section, let us address the reason why we can reproduce only a pair of solutions by the Born approximation.
We show that the pair of solutions~(\ref{eq3-130}) according to the Born approximation actually corresponds to the pair of the case $n=1$ of the exact solutions~(\ref{eq3-1040}).
The exact solutions with $n\geq2$ emerge because of higher-order effects.

For $V_0\ll{\kappa_0}^2$, the Bessel function in Eq.~(\ref{eq3-1030}) for odd solutions can be expanded in the form
\begin{align}\label{eq3-140}
0=J_\nu(2\alpha)=\alpha^{\nu/2}\sum_{n=0}^\infty\frac{(-1)^n\alpha^{n}}{n!\Gamma(\nu+n+1)},
\end{align}
where $\nu=-2ik_\textrm{res}/\kappa_0$ and $\alpha=V_0/{\kappa_0}^2$.
Suppose that we terminate the summation in Eq.~(\ref{eq3-140}) at $n=1$, that is, at the order $O(V_0)$, and assume the deviation of $\nu$ from the value $-1$ as $\nu=-1+\delta_1$.
The resonance equation~(\ref{eq3-140}) then reduces to
\begin{align}\label{eq3-150}
0=\frac{1}{\Gamma(\delta_1)}-\frac{\alpha}{\Gamma(\delta_1+1)}=\frac{1}{\Gamma(\delta_1)}\left(1-\frac{\alpha}{\delta_1}\right).
\end{align}
This is followed by $\delta_1=\nu+1=\alpha$, which is indeed the solution~(\ref{eq3-110}) by the Born approximation.
Next we terminate the summation in Eq.~(\ref{eq3-140}) at $n=2$, or at the order $O({V_0}^2)$, and assume $\nu=-2+\delta_2$.
The resonance equation~(\ref{eq3-140}) then reduces to
\begin{align}\label{eq3-160}
0=\frac{1}{\Gamma(-1+\delta_2)}-\frac{\alpha}{\Gamma(\delta_2)}+\frac{\alpha^2}{2\Gamma(1+\delta_2)}
=\frac{1}{\Gamma(\delta_2)}\left[
(-1+\delta_2)-\alpha+\frac{\alpha^2}{2\delta_2}
\right].
\end{align}
We numerically confirmed that the solution of Eq.~(\ref{eq3-160}) well reproduces the exact solution~(\ref{eq3-1040}) with $n=2$ for $\alpha\ll1$.
We went further and confirmed this correspondence up to $n=5$.
This shows that the exact solutions~(\ref{eq3-1040}) with $n\geq2$ are indeed higher-order effects.

Summarizing the above discussion, we stress the following point.
It is not a coincidence that the Born approximation reproduced the resonant solutions on the negative imaginary axis.
On the contrary, the Born approximation successfully reproduced the first pair of the anti-bound states.

\section{An exponential potential with an oscillation}
\label{sec4}

In the present section, we discuss the resonance spectrum for an exponential potential with an oscillation;
\begin{align}\label{eq4-10}
V(x)=-V_0\cos(\omega_0 x)e^{-\kappa_0|x|},
\end{align}
where $V_0>0$ and $\kappa_0>0$.
There is no exact solution for this potential.
We discuss this case as a primer to the case in the next section, where we discuss an exponential potential with a cutoff.
We will argue in the next section that a cutoff plays a role of superposition of various oscillations.
In the present section, we are going to show that an oscillatory factor in Eq.~(\ref{eq4-10}) makes each of the anti-bound states~(\ref{eq3-120}) split into a pair of resonant and anti-resonant states.
Then in the next section, we will show that a cutoff, because of its multiple oscillatory Fourier components, yields a \textit{sequence} of pairs of resonant and anti-resonant states.

We can rewrite the potential~(\ref{eq4-10}) as a sum of two exponential functions~\cite{Martin59};
\begin{align}\label{eq4-20}
V(x)=-\frac{V_0}{2}\left[e^{-(\kappa_0+i\omega_0)|x|}+e^{-(\kappa_0-i\omega_0)|x|}\right].
\end{align}
By defining
\begin{align}\label{eq4-30}
\kappa_1\equiv\kappa_0+i\omega_0,
\quad
\kappa_2\equiv\kappa_0-i\omega_0,
\quad
V_1=V_2\equiv\frac{V_0}{2},
\end{align}
we can reduce the potential~(\ref{eq4-20}) into the form
\begin{align}\label{eq4-40}
V(x)=-\sum_{j=1}^2V_je^{-\kappa_j|x|}.
\end{align}
Hence the resonance equation is given by
\begin{align}\label{eq4-50}
1+\sum_{j=1}^{2}\frac{V_j}{2ik_\textrm{res}}\left(\frac{1}{\kappa_j}+\sigma\frac{1}{\kappa_j-2ik_\textrm{res}}\right)=0.
\end{align}



The solutions within the Born approximation for multiple exponentials are given by Eq.~(\ref{eq3-130}).
In the present case of Eq.~(\ref{eq4-30}), we readily obtain the solutions as
\begin{align}\label{eq4-100}
k_\textrm{res}&=
-\frac{i}{2}\left(\kappa_0\pm i\omega_0+\sigma\frac{V_0}{2}\frac{1}{\kappa_0\pm i\omega_0}\right)
\nonumber\\
&=\pm\frac{1}{2}\omega_0\left(1-\sigma\frac{V_0/2}{{\kappa_0}^2+{\omega_0}^2}\right)
-\frac{i}{2}\kappa_0\left(1+\sigma\frac{V_0/2}{{\kappa_0}^2+{\omega_0}^2}\right).
\end{align}
The answer converges to~(\ref{eq3-120}) as $\omega_0\to0$;
note that we should also replace $V_0/2$ with $V_0$ to make the exact correspondence.
The largest difference between Eq.~(\ref{eq3-120}) and Eq.~(\ref{eq4-100}) is the existence of the real part.
Each of the anti-bound states given by Eq.~(\ref{eq3-120}) now splits into a pair of a resonant state and an anti-resonant state and shifts sideways.
If we have more than one exponentials with different oscillations, each anti-bound state splits into a pair as illustrated in Fig.~\ref{fig1}(b).

\section{An exponential potential with a cutoff}
\label{sec5}

\begin{figure}[t]
\centering
\includegraphics[width=0.7\textwidth]{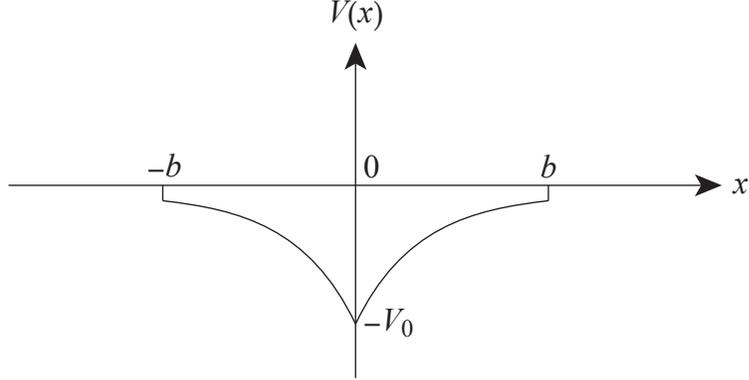}
\caption{An exponential potential with a cutoff.}
\label{fig2}
\end{figure}

In the present section, we discuss the resonance spectrum for an exponential potential with a cutoff (Fig.~\ref{fig2});
\begin{align}\label{eq5-10}
V(x)=
\begin{cases}
-V_0e^{-\kappa_0|x|} & \mbox{for $|x|\leq b$,} \\
0 & \mbox{for $|x| > b$,}
\end{cases}
\end{align}
where $V_0>0$, $\kappa_0>0$ and $b>0$.
This potential is exactly solvable;
we nonetheless treat the problem in the present framework of the Born approximation.
The purpose is to see explicitly the origin of a resonance spectrum surprisingly different from the case of a simple exponential potential.

As we described in section~\ref{sec3}, an exponential potential produces only anti-bound states.
In striking contrast, an exponential potential with a cutoff produces a series of extra resonant states below and along the real axis of the complex wave-number plane.
Let us start with somewhat hand-waving argument, where we attribute this structure to an exponential with various oscillations.
The discontinuity at the cutoffs $x=\pm b$ may be described as superposition of various Fourier components.
In other words, the cutoff potential~(\ref{eq5-10}) may be written in the form
\begin{align}\label{eq5-20}
V(x)\sim-V_0e^{-\kappa_0|x|}\sum_n c_n \cos\omega_nx.
\end{align}
If we accept this, Eq.~(\ref{eq4-100}) dictates that we have a series of resonant poles at
\begin{align}\label{eq5-30}
k_\textrm{res}\simeq \frac{\pm \omega_n-i\kappa_0}{2}.
\end{align}
These poles are aligned below and along the real $k$ axis.

For a little bit more detailed analysis, let us express the potential~(\ref{eq5-10}) as
\begin{align}\label{eq5-40}
V(x)=-V_0e^{-\kappa_0|x|}\times\left(\Theta(x+b)-\Theta(x-b)\right),
\end{align}
where $\Theta(x)$ is the Heaviside step function.
The Fourier expansion of the term with the step functions is given by
\begin{align}\label{eq5-50}
\Theta(x+b)-\Theta(x-b)=2\int_{-\infty}^\infty \frac{\sin kb}{k}e^{ikx}dk.
\end{align}
The Fourier coefficient $\sin kb/k$ has peaks and dips at $k=(2n-1)\pi/(2b)$ with an integer $n$ except for $k=\pm \pi/(2b)$.
Thus we could argue that the approximation~(\ref{eq5-20}) is legitimate with
\begin{align}\label{eq5-55}
\omega_n=\frac{2n-1}{2b}\pi
\end{align}
for large $|n|$.
Indeed, we will see in Eq.~(\ref{eq5-170}) below that the location of the poles can be approximated as
\begin{align}\label{eq5-60}
k_\textrm{res}\simeq \frac{(2n-1)\pi}{4b}-i\frac{\kappa_0}{2}
\end{align}
for large $|n|$.
This is consistent with Eq.~(\ref{eq5-30}) accompanied by Eq.~(\ref{eq5-55}).

Let us now develop the Born approximation for the exponential potential with a cutoff, Eq.~(\ref{eq5-10}).
The formula~(\ref{eq2-60}) gives
\begin{align}\label{eq5-70}
\lim_{x\to+\infty}\phi(k;x)
&=\sigma e^{-ikx}+e^{ikx}\left\{1-\frac{V_0}{2ik}\left[
\frac{2}{\kappa_0}+\frac{\sigma}{\kappa_0-2ik}+\frac{\sigma}{\kappa_0+2ik}\right.\right.
\nonumber\\
&\phantom{=\sigma e^{-ikx}+e^{ikx}\Bigg\{1}\left.\left.
-e^{-\kappa_0 b}\left(
\frac{2}{\kappa_0}+\frac{\sigma e^{2ikb}}{\kappa_0-2ik}+\frac{\sigma e^{-2ikb}}{\kappa_0+2ik}
\right)\right]\right\}.
\end{align}
After the procedure given in Eq.~(\ref{eq3-30}), we have the resonance equation as
\begin{align}\label{eq5-80}
0&=1+\frac{V_0}{2ik_\textrm{res}}\left[
\frac{1}{\kappa_0}+\frac{\sigma}{\kappa_0-2ik_\textrm{res}}
-e^{-\kappa_0 b}\left(
\frac{1}{\kappa_0}+\frac{\sigma e^{2ik_\textrm{res}b}}{\kappa_0-2ik_\textrm{res}}
\right)\right]
\nonumber\\
&=1+\frac{V_0}{2ik_\textrm{res}}\left[\frac{1-e^{-\kappa_0 b}}{\kappa_0}
+\sigma\frac{1-e^{-(\kappa_0-2ik_\textrm{res})b}}{\kappa_0-2ik_\textrm{res}}\right]
\end{align}

In order to solve this resonance equation, we again note that the dominant contribution comes from the term with the denominator $\kappa_0-2ik_\textrm{res}$.
We therefore have
\begin{align}\label{eq5-90}
2ik_\textrm{res}(2ik_\textrm{res}-\kappa_0)
=\sigma V_0\left[1-e^{(2ik_\textrm{res}-\kappa_0)b}\right].
\end{align}
For the equation to converge in the limit $b\to\infty$, the real part in the exponent must not depend on $b$.
We therefore assume the form
\begin{align}\label{eq5-100}
2ik_\textrm{res}=\kappa_0+\frac{\alpha+i\beta}{b},
\end{align}
where $\alpha$ and $\beta$ are real.
Using the form~(\ref{eq5-100}) in Eq.~(\ref{eq5-90}) for large $b$, we have
\begin{align}\label{eq5-110}
\alpha+i\beta
=\sigma \frac{V_0 b}{\kappa_0}\left(1-e^{\alpha+i\beta}\right),
\end{align}
or
\begin{align}\label{eq5-120}
\alpha-\sigma\frac{V_0 b}{\kappa_0}=-\sigma\frac{V_0 b}{\kappa_0}e^\alpha\cos\beta,
\qquad
\beta=-\sigma\frac{V_0 b}{\kappa_0}e^\alpha\sin\beta.
\end{align}
Taking the ratio of the two, we obtain
\begin{align}\label{eq5-130}
\tan\beta=\frac{\beta}{\alpha-\sigma V_0 b/\kappa_0}.
\end{align}
whose solution is approximated for large $|\beta|$ by
\begin{align}\label{eq5-140}
\beta=\frac{2n-1}{2}\pi+\delta_n,
\end{align}
where $n$ is an integer and $\delta_n$ denotes the correction.
Equation~(\ref{eq5-120}) is then approximated by
\begin{align}\label{eq5-150}
\alpha-\sigma\frac{V_0 b}{\kappa_0}&\simeq\pm \sigma\frac{V_0 b}{\kappa_0}e^\alpha \delta_n,
\\ \label{eq5-151}
-\frac{2n-1}{2}\pi&\simeq\pm\sigma\frac{V_0 b}{\kappa_0}e^\alpha,
\end{align}
where, on the right-hand sides, we choose the positive signs for odd $n$ and the negative signs for even $n$.
(Since the left-hand side of Eq.~(\ref{eq5-151}) is negative for $\beta>0$, we realize that in the region $\mathop{\textrm{Re}}k_\textrm{res}>0$, the symmetric solutions $\sigma=1$ appear for even $n$ and the antisymmetric solutions $\sigma=-1$ appear for odd $n$.
The situation is the opposite for $\mathop{\textrm{Re}}k_\textrm{res}<0$.)
We then have
\begin{align}\label{eq5-160}
\alpha&=\ln\left(\frac{\kappa_0}{V_0 b}\frac{|2n-1|}{2}\pi\right),
\\ \label{eq5-161}
\delta_n&=\frac{2}{(2n-1)\pi}\left[\sigma\frac{V_0 b}{\kappa_0}-\ln\left(\frac{\kappa_0}{V_0 b}\frac{|2n-1|}{2}\pi\right)\right].
\end{align}
To summarize, we arrive at the solutions
\begin{align}\label{eq5-170}
k_\textrm{res}&=\frac{2n-1}{4b}\pi
+\frac{1}{(2n-1)\pi}\left[\sigma\frac{V_0}{\kappa_0}-\frac{1}{b}\ln\left(\frac{\kappa_0}{V_0 b}\frac{|2n-1|}{2}\pi\right)\right]
\nonumber\\
&\phantom{=}-\frac{i}{2}\left[\kappa_0+\frac{1}{b}\ln\left(\frac{\kappa_0}{V_0 b}\frac{|2n-1|}{2}\pi\right)\right]
\end{align}
for large $b$ and $n$, or, in the lowest order, Eq.~(\ref{eq5-60}).
As we emphasized above, this is consistent with the solutions for the exponential potential with oscillations generated by the two step functions~(\ref{eq5-40}).
In short, the introduction of a cutoff possibly for numerical calculations can be disastrous, generating infinitely many extra poles that would not exist without the cutoff.

\begin{figure}
\centering
\includegraphics[width=0.7\textwidth]{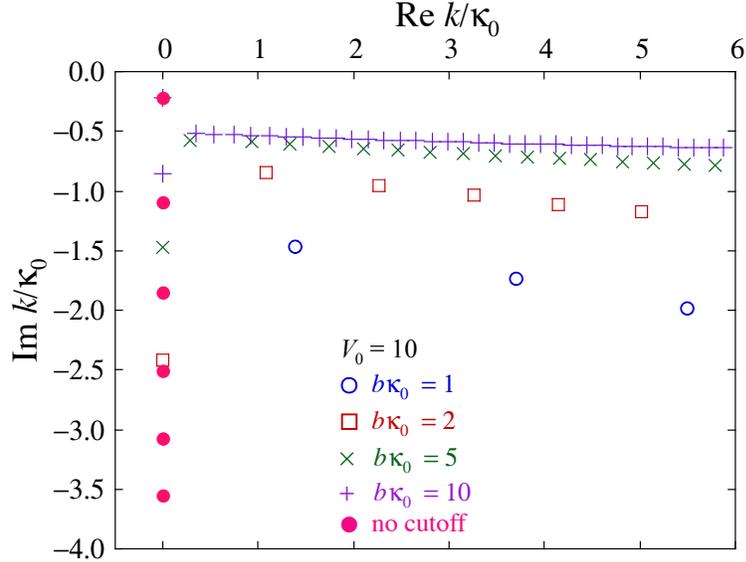}
\caption{Exact resonant solutions of an exponential potential with a cutoff with $V_0=10$ and $b=1,2,5,10[/\kappa_0]$.}
\label{fig3}
\end{figure}

\begin{figure}
\centering
\includegraphics[width=0.8\textwidth]{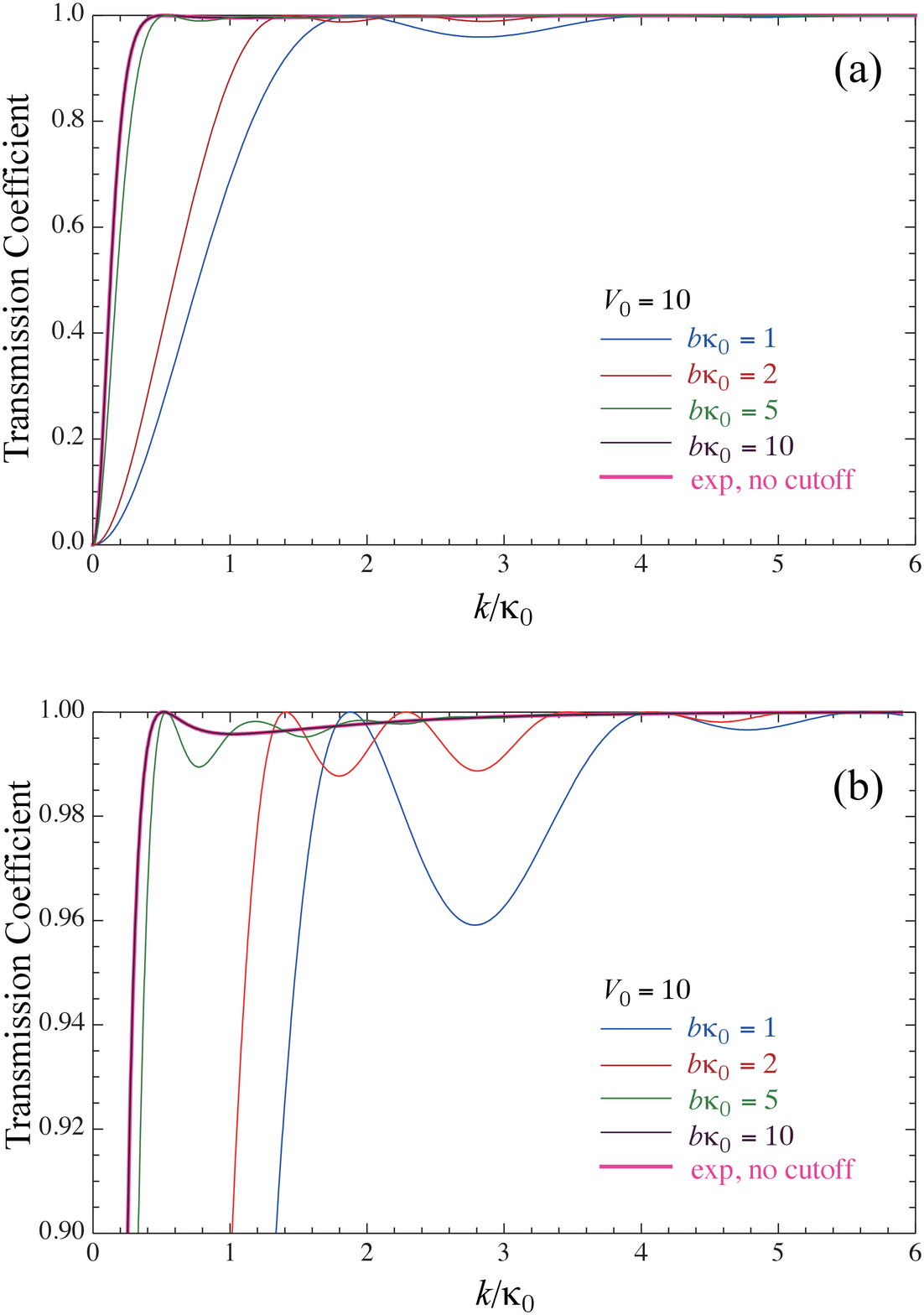}
\caption{The transmission coefficient of the exponential potential with a cutoff with $V_0=10$ and $b=1,2,5,10[/\kappa_0]$ (from the right curve to the left curve in this order). The panel (b) is a blowup of the panel (a). The case $b\kappa_0=10$ is indiscernible from the case without a cutoff.}
\label{fig4}
\end{figure}

At the end of the present section, we show in Fig.~\ref{fig3}, the results of the exact solution for the exponential potential with a cutoff, Eq.~(\ref{eq5-10}), and compare them with the result of the exact solution for the exponential potential without a cutoff, Eq.~(\ref{eq3-10}).
The results are consistent with the Born approximation~(\ref{eq5-170}) particularly for large $b$.
The series of poles below and along the real $k$ axis become denser and denser as we increase the cutoff range $b$.
In fact, the residues of the poles decrease exponentially as the discontinuity at the cutoff $\exp(-\kappa_0 b)$ decreases.
Since the density of the poles increases only proportionally to $b$, the whole contributions of the resonant and anti-resonant poles vanish in the limit $b\to\infty$ and hence only the anti-bound states remain for the exponential potential without a cutoff.
This is also observed in the transmission coefficient shown in Fig.~\ref{fig4}.
The resonance peaks (or dips) become less prominent as we increase $b$ and eventually indiscernible from the case without a cutoff.

\section{Gaussian potential}
\label{sec6}

In the present section, we discuss the resonance spectrum for a potential with one or multiple Gaussian functions.
We start with a potential with one Gaussian function
\begin{align}\label{eq6-10}
V(x)=-V_0 e^{-{a_0}^2x^2},
\end{align}
where $V_0>0$ and $a_0>0$.
We will show that the series of resonant eigen-wave-numbers asymptotically approaches the line $\mathop{\textrm{Im}} k=-\mathop{\textrm{Re}} k$ from below, or the line of argument $-\pi/4$ in the complex $k$ plane from below, and the imaginary axis $\mathop{\textrm{Re}} E=0$ from left.

The formula~(\ref{eq2-60}) for the Born approximation in this case yields
\begin{align}\label{eq6-20}
\lim_{x\to+\infty}\phi(k;x)=
\sigma e^{-ikx}+e^{ikx}
\left[
1-\frac{V_0}{2ik}
\frac{\sqrt{\pi}}{a_0}
\left(1+\sigma e^{-k^2/{a_0}^2}\right)
\right].
\end{align}
By changing the normalization, we have
\begin{align}\label{eq6-30}
\lim_{x\to+\infty}\phi(k;x)\propto
e^{ikx}+\sigma e^{-ikx}\left[
1+\frac{V_0}{2ik}
\frac{\sqrt{\pi}}{a_0}
\left(1+\sigma e^{-k^2/{a_0}^2}\right)
\right],
\end{align}
which is followed by the resonance equation
\begin{align}\label{eq6-40}
2ik_\textrm{res}+
\frac{V_0\sqrt{\pi}}{a_0}
\left(1+\sigma e^{-{k_\textrm{res}}^2/{a_0}^2}\right)=0.
\end{align}
Within the Born approximation, the resonance equation for the potential with multiple Gaussians
\begin{align}\label{eq6-50}
V(x)=-\sum_j V_j e^{-{a_j}^2 x^2}
\end{align}
is given by
\begin{align}\label{eq6-60}
2ik_\textrm{res}+\sum_j
\frac{V_j\sqrt{\pi}}{a_j}
\left(1+\sigma e^{-{k_\textrm{res}}^2/{a_j}^2}\right)=0.
\end{align}

Let us now solve the resonance equation~(\ref{eq6-40}) for large $|k_\textrm{res}|$.
For the resonance equation~(\ref{eq6-40}) to be stable, the real part of the exponent ${k_\textrm{res}}^2/{a_0}^2$ should stay small
\begin{align}\label{eq6-65}
\mathop{\textrm{Re}}\frac{{k_\textrm{res}}^2}{{a_0}^2}=O(1)
\end{align}
for large $|k_\textrm{res}|$.
This means that if we assume the form
\begin{align}\label{eq6-70}
\frac{k_\textrm{res}}{a_0}=\alpha-i\beta
\end{align}
with positive $\alpha$ and $\beta$, we should have
$\alpha-\beta=O(1)$.
This is then followed by $2ik_\textrm{res}\propto 1+i$.
Equation~(\ref{eq6-40}) then gives
\begin{align}\label{eq6-80}
\exp\left[-\frac{{k_\textrm{res}}^2}{{a_0}^2}\right]\propto -\sigma(1+i)
\end{align}
with a positive proportional coefficient, or
\begin{align}\label{eq6-90}
-\mathop{\textrm{Im}}\frac{{k_\textrm{res}}^2}{{a_0}^2}=2\alpha\beta\simeq\left(n-\frac{3}{4}\right)\pi
\end{align}
for a large integer $n$.
We choose an even $n$ for a symmetric solution $\sigma=+1$ and an odd $n$ for an antisymmetric solution $\sigma=-1$.
Let us then denote the solutions of $\alpha$ and $\beta$ as $\alpha_n$ and $\beta_n$, respectively, for the integer $n$.
We can assume the following:
\begin{align}\label{eq6-100}
{\alpha_n}^2-{\beta_n}^2&=-\gamma_n,
\\ \label{eq6-105}
\alpha_n\simeq \beta_n &\simeq \sqrt{\frac{\pi}{2}}\sqrt{n},
\end{align}
where $\gamma_n$ is a real number of order $n^0$.
We will then ask for the behavior of $\gamma_n$.

Equations~(\ref{eq6-100}) and~(\ref{eq6-105}) give
\begin{align}\label{eq6-110}
\exp\left[-\left(\frac{k_\textrm{res}}{a_0}\right)^2\right]
=\exp\left[\gamma_n+\left(n-\frac{3}{4}\right)\pi i\right]=-\frac{\sigma}{\sqrt{2}}(1+i)e^{\gamma_n},
\end{align}
where on the right-hand side we use the fact that $\sigma=+1$ for even $n$ and $\sigma=-1$ for odd $n$.
The resonance equation~(\ref{eq6-40}) now reduces to
\begin{align}\label{eq6-120}
2\sqrt{n}-\frac{V_0}{{a_0}^2}e^{\gamma_n}\simeq 0.
\end{align}
We thus have the asymptotics
\begin{align}\label{eq6-130}
\gamma_n\simeq\frac{1}{2}\ln n.
\end{align}
Note that the asymptotics~(\ref{eq6-130}) is compatible with the asymptotics~(\ref{eq6-105}), because $\ln n$ is of lower order of $n$ than $\sqrt{n}$.

\begin{figure}
\begin{minipage}[b]{0.48\textwidth}
\includegraphics[width=\textwidth]{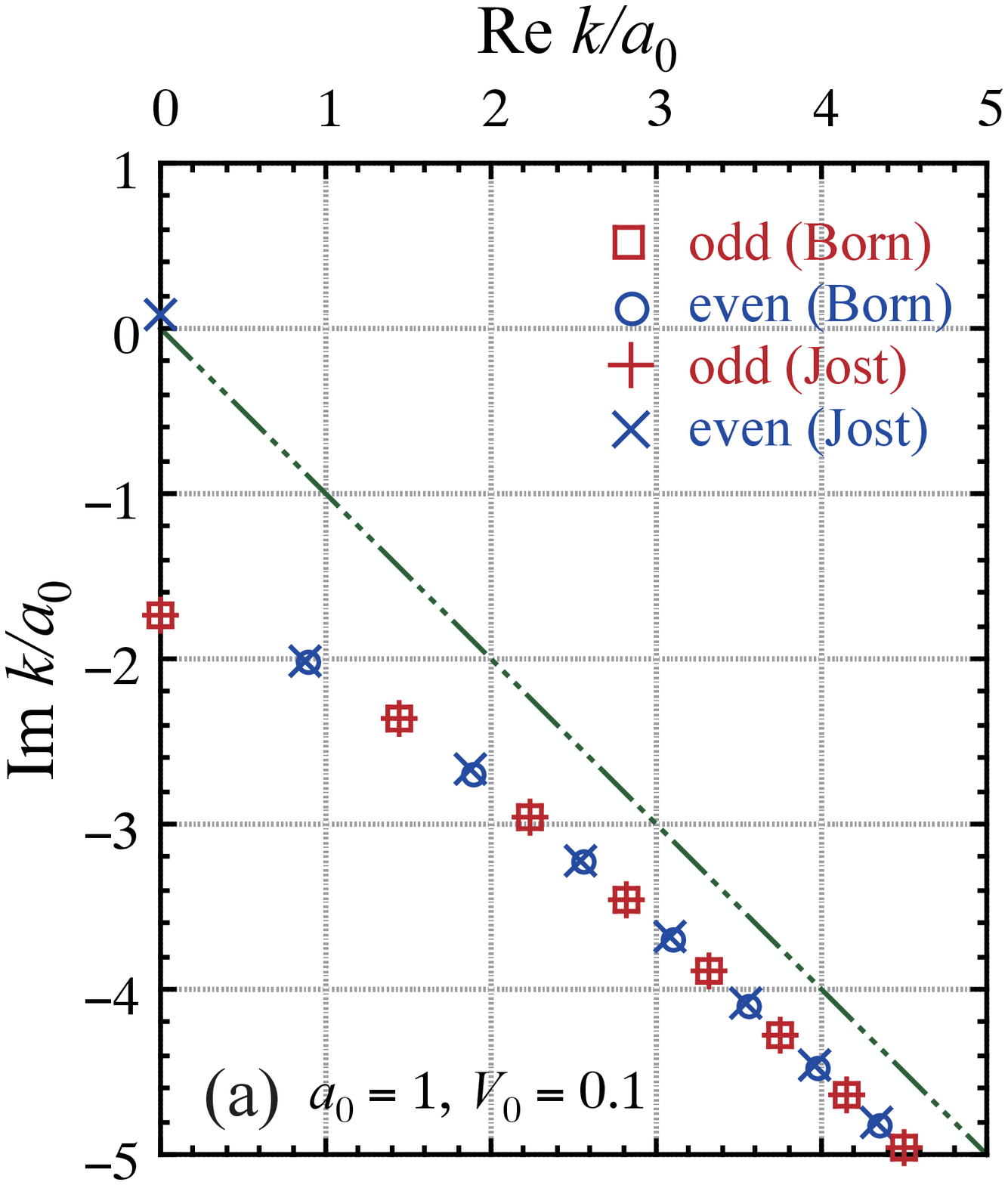}
\vspace{0mm}
\end{minipage}
\hfill
\begin{minipage}[b]{0.48\textwidth}
\includegraphics[width=\textwidth]{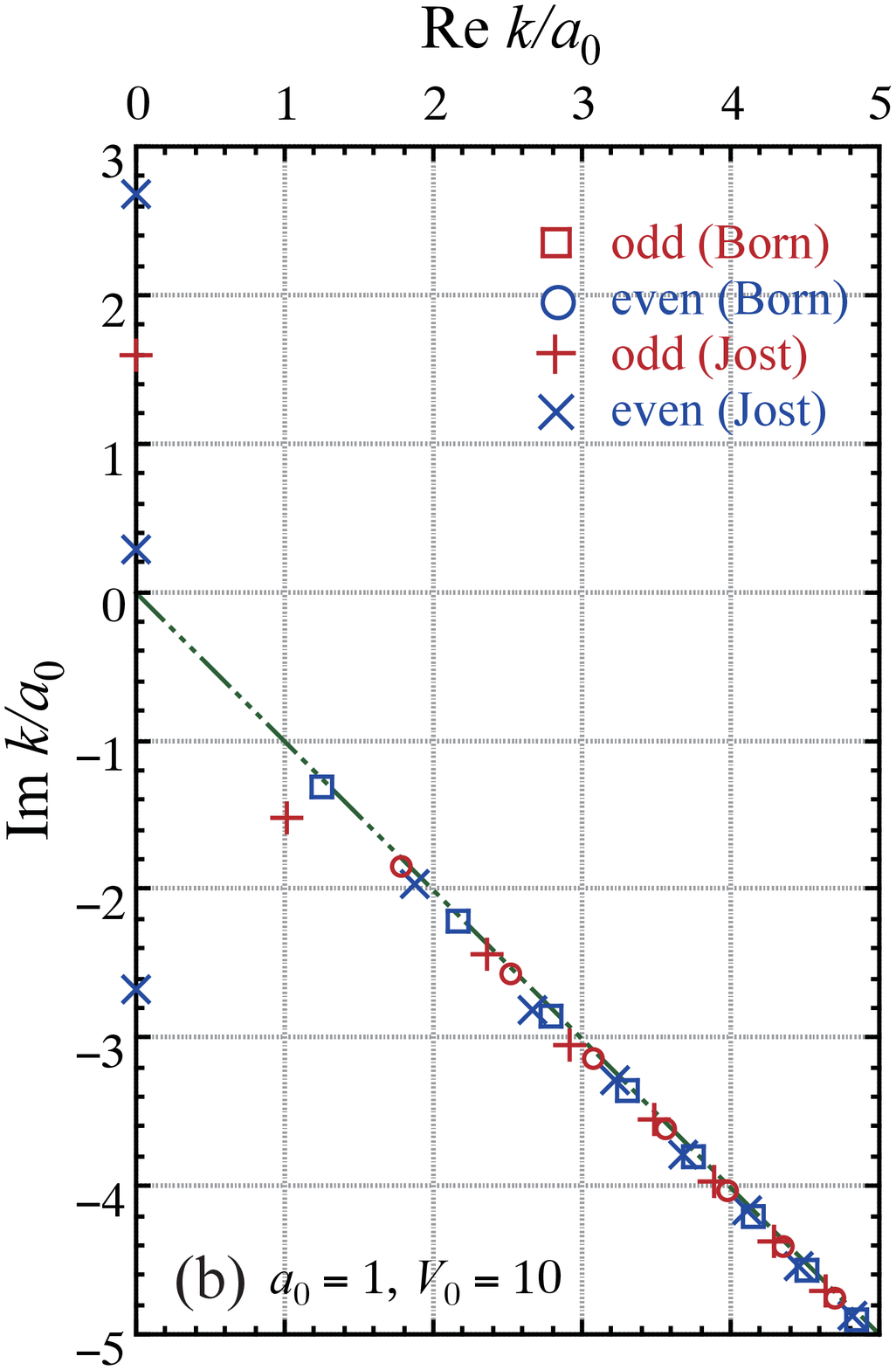}
\vspace{0mm}
\end{minipage}
\caption{The distribution of resonance poles in the complex wave-number plane obtained by the Born approximation and the Jost-function method.
(a) $a_0=1$ and $V_0=0.1$ and (b) $a_0=1$ and $V_0=10$.
In the latter case, we have three bound states.}
\label{fig5}
\end{figure}

We compare in Fig.~\ref{fig5} the results of the Born approximation with the results of the Jost-function method, which is described in Appendix B.
The Jost-function method is supposedly producing numerically exact solutions here.
We can see that the results of the Born approximation are almost equal to the numerically exact solutions for $V_0=0.1$ and fairly consistent with them in the region $\mathop{\textrm{Re}}k/a_0\gtrsim 3$ for large $V_0=10$.

The resonant poles approach the line of argument $-\pi/4$ from below in the complex $k$ plane, which is also consistent with the result of the Born approximation, $\gamma_n>0$.
In the case of the strong potential, $V_0=10$, the resonant poles are much closer to the $-\pi/4$ line, but they still approach the line from below.
This means that all the resonant eigenenergies have negative real parts:
\begin{align}\label{eq6-140}
\mathop{\textrm{Re}}E_\textrm{res}=\mathop{\textrm{Re}}{k_\textrm{res}}^2<0.
\end{align}

Finally, for a positive potential ($V_0<0$ in the present notation), Eq.~(\ref{eq6-110}) reads
\begin{align}\label{eq6-150}
\exp\left[-\left(\frac{k_\textrm{res}}{a_0}\right)^2\right]
=\exp\left[\gamma_n+\left(n+\frac{1}{4}\right)\pi i\right]=\frac{\sigma}{\sqrt{2}}(1+i)e^{\gamma_n}
\end{align}
with $\sigma=+1$ for even $n$ and $\sigma=-1$ for odd $n$.
In other words, the even solutions and odd solutions are interchanged with each other.
Equation~(\ref{eq6-120}) reads
\begin{align}\label{eq6-160}
2\sqrt{n}-\frac{|V_0|}{{a_0}^2}e^{\gamma_n}\simeq 0
\end{align}
with the same Eq.~(\ref{eq6-130}).
The fact that the resonant poles approach the line of arugmet $-\pi/4$ from below remains the same.

\section{Summary and discussions}

In the present paper, we formulated the Born approximation for finding resonance poles in the complex plane.
We found it working well for scattering potentials of exponential functions as well as Gaussians.

The greatest merit of using the Born approximation is that we can see the origin of the structure of the resonance spectra.
In the case of the exponential potential with the cutoff in Sec.~\ref{sec5}, we were able to see clearly that the reason why the extra resonance poles are aligned below and along the real $k$ axis is because the cutoff generates oscillatory potentials.

The hand-waiving argument around and below Eq.~(\ref{eq5-20}), which was confirmed by the Born approximation, should be applicable to any scattering potentials with discontinuities.
We thereby propose the following conjecture:

\vspace{\baselineskip}
\noindent
\hspace{10mm}
\begin{minipage}[t]{120mm}
\textbf{Conjecture 1:} Any discontinuities in the scattering potential cause a series of extra resonance poles below and along the real $k$ axis as well as along the real $E$ axis.
\end{minipage}
\vspace{\baselineskip}

We were also able to explain clearly the reason why the resonance poles approach the $-\pi/4$ line in the complex $k$ plane from below for the Gaussian potential.
The reason of the convergence to the $-\pi/4$ line in the complex $k$ plane is the fact $\alpha-\beta=O(1)$, which follows from Eq.~(\ref{eq6-65}).
We note that the square of $k_\textrm{res}$ in Eq.~(\ref{eq6-65}), or in the exponent of the resonance equation~(\ref{eq6-40}), is originated from the square of $x$ in the exponent of the Gaussian potential~(\ref{eq6-10}).
It is then reasonable to conjecture the following;

\vspace{\baselineskip}
\noindent
\hspace{10mm}
\begin{minipage}[t]{120mm}
\textbf{Conjecture 2:} For a scattering potential of the form
\begin{align}\label{eq7-10}
V(x)=-V_0e^{-{a_0}^\lambda |x|^\lambda},
\end{align}
the resonance poles should approach the line of the argument $-\pi/(2\lambda)$ in the complex $k$ plane and the line of the argument $-\pi/\lambda$ in the complex $E$ plane.
\end{minipage}
\vspace{\baselineskip}

As we saw in Sec.~\ref{sec3}, we indeed have poles only on the imaginary $k$ axis for $\lambda=1$, namely for the exponential potential.
We could also regard that the limit $\lambda\to\infty$ corresponds to a potential with a very rapid decrease, or a cutoff, in which case the poles are indeed aligned along the real $k$ axis.

\section*{Acknowledgements}
The present author is grateful to Prof.~K.~Kat\={o} and Prof.~N.~Moiseyev for interesting discussion and helpful encouragement.
This work is supported by Core Research for Evolutional Science and  Technology (CREST) of Japan Science and Technology Agency.

\appendix

\section{Jost solutions and Jost functions in one dimension}

In this section, we derive the Jost solutions and the Jost functions in one dimension.
In the usual context, they are derived in the radial coordinate of the three-dimensional space, which is a semi-infinite one-dimensional space.
We here consider the one-dimensional space which is infinite in both positive and negative directions.
We hence suppose that reviewing the derivation for the one-dimensional space may be instructive.

When the potential is symmetric with respect to the origin, any solutions can be classified into even or odd solutions.
In fact, the solutions in the radial coordinate are equivalent to odd solutions in the infinite one-dimensional space, because they must vanish at the origin.

We define the Jost solutions $f_\pm(k;x)$ by their behavior in the far right:
\begin{align}\label{eqa-10}
\lim_{x\to\infty}e^{\mp ikx}f_\pm(k;x)=1.
\end{align}
For a symmetric potential, the Jost functions are also classified to even and odd solutions.
Therefore, the Jost solutions behave in the far left as
\begin{align}\label{eqa-20}
\lim_{x\to-\infty}e^{\pm ikx}f_\pm(k;x)=\sigma
\equiv\begin{cases}
+1 & \mbox{for odd solutions,}\\
-1 & \mbox{for even solutions.}
\end{cases}
\end{align}

The Jost solutions are generally discontinuous or undifferentiable at the origin.
On the other hand, a physical solution $\phi(k;x)$ must be continuous and differentiable at the origin.
The boundary conditions for the physical solution at the origin can be given in the forms
\begin{align}\label{eqa-30}
\phi(k;0)=1 \quad\mbox{and}\quad \phi'(k;0)=0 & \quad\mbox{for even solutions,}
\\ \label{eqa-31}
\phi(k;0)=0 \quad\mbox{and}\quad \phi'(k;0)=1 & \quad\mbox{for odd solutions,}
\end{align}
where $\phi'(k;x)=\partial \phi(k;x)/\partial x$.

Since the Jost solutions $f_\pm(k;x)$ are mutually independent solutions of a second differential equation, we can express the physical solution as a linear combination of the Jost solutions:
\begin{align}\label{eqa-40}
\phi(k;x)=C_+(k)f_+(k;x)+C_-(k)f_-(k;x).
\end{align}
In order to proceed, we use the Wronskian
\begin{align}\label{eqa-1010}
W(f,g)\equiv f(x)g'(x)-g(x)f'(x),
\end{align}
which must be independent of $x$ for linearly independent solutions $f$ and $g$ of the Schr\"{o}dinger equation~(\ref{eq1-5}).
Since we can obtain $W(f_+,f_-)=-2ik$ in the limit $x\to\infty$, we have
\begin{align}\label{eqa-1020}
W(f_+,\phi)&=C_-(k)W(f_+,f_-)=-2ikC_-(k),
\\ \label{eqa-1021}
W(f_-,\phi)&=C_+(k)W(f_-,f_+)=2ikC_+(k)
\end{align}
in the limit $x\to\infty$.
In the limit $x\to 0$, on the other hand, we have
\begin{align}\label{eqa-50}
F_\pm(k)\equiv W(f_\pm,\phi)&=f_\pm(k;0)\phi'(k;0)-\phi(k;0)f_\pm'(k;0)
\nonumber\\
&=\begin{cases}
-f_\pm'(k;0) &\quad\mbox{for even solutions},\\
f_\pm(k;0) &\quad\mbox{for odd solutions,}
\end{cases}
\end{align}
where we used the boundary conditions~(\ref{eqa-30}) and~(\ref{eqa-31}) in the second equation.
The functions $F_\pm(k)$ defined above are called the Jost functions, which should not be confused with the Jost solutions $f_\pm(k;x)$.
(The Jost function for even solutions may not be common because most papers discuss only radial functions in three dimensions, in which case the solutions must vanish at the origin.)
Equating Eqs.~(\ref{eqa-1020})--(\ref{eqa-50}), we obtain $C_\pm(k)=\pm F_\mp(k)/(2ik)$.
The physical solution~(\ref{eqa-40}) is now given in the form
\begin{align}\label{eqa-120}
\phi(k;x)=\frac{1}{2ik}\left(F_-(k)f_+(k;x)-F_+(k)f_-(k;x)\right).
\end{align}

The Siegert condition for a resonance pole is
\begin{align}\label{eqa-140}
\lim_{|x|\to\infty}\phi(k_\mathrm{res};x)e^{-ik_\mathrm{res}|x|}=\mbox{constant}.
\end{align}
In other words, we should have only the first term on the left-hand side of Eq.~(\ref{eqa-120}).
Therefore, the resonance equation is given by the zeros of the Jost function,
\begin{align}\label{eqa-150}
F_+(k_\mathrm{res})=0.
\end{align}

\section{Jost-function method of finding resonant states}

In the present section, we describe some details of the Jost-function method, which we used to obtain the results in Fig.~\ref{fig5}.
The method was proposed in Refs.~\citen{Rakityansky96,Sofianos97,Rakityansky98} and was extended in Ref.~\citen{Masui99}.
They all obtained the solutions for Coulombic scattering potentials with an angular momentum.
Since we do not include a Coulomb potential nor an angular momentum in the present paper, the formulation becomes simpler than the one described in the references.
We consider that it may be worth reviewing here the simpler version.

Let us find the solutions of the Schr\"{o}dinger equation~(\ref{eq1-5}) of a form other than Eq.~(\ref{eqa-120}).
If we did not have the potential $V(x)$, the solution would be given by a linear combination of the simple plane waves $e^{\pm ikx}$,
\begin{align}\label{eqb-160}
\phi(k;x)=\frac{1}{2ik}\left(F_-(k)e^{ikx}-F_+(k)e^{-ikx}\right)
\quad\mbox{for $V(x)\equiv0$}
\end{align}
with the Jost functions $F_\pm(k)$ for $V(x)=0$ being the coefficients.
When we turn on the scattering potential $V(x)$, we can write down a solution by giving the two coefficients the $x$-dependence:
\begin{align}\label{eqb-165}
\phi(k;x)=\frac{1}{2ik}\left(\mathcal{F}_-(k;x)e^{ikx}-\mathcal{F}_+(k;x)e^{-ikx}\right)
\quad\mbox{for $V(x)\neq0$}.
\end{align}
That is, instead of modifying the plane waves $e^{\pm ikx}$ to the Jost solutions $f_\pm(k;x)$ as in Eq.~(\ref{eqa-120}), we modify the coefficients from $F_\pm(k)$ to $\mathcal{F}_\pm(k;x)$.
By comparing Eqs.~(\ref{eqa-120}) and~(\ref{eqb-165}), we notice
\begin{align}\label{eqb-167}
\lim_{x\to\infty}\mathcal{F}_\pm(k;x)=F_\pm(k).
\end{align}
Therefore, the resonance equation~(\ref{eqa-150}) is now converted into
\begin{align}\label{eqb-168}
\lim_{x\to\infty}\mathcal{F}_+(k_\mathrm{res};x)=0.
\end{align}

The remaining task for us is to derive differential equations for the functions $\mathcal{F}_\pm(k;x)$.
There is, in fact, redundancy in the functions $\mathcal{F}_\pm(k;x)$, something like a gauge degree of freedom.
In order to fix the gauge, we assume a condition
\begin{align}\label{eqb-170}
\mathcal{F}'_-(k;x)e^{ikx}-\mathcal{F}'_+(k;x)e^{-ikx}\equiv0,
\end{align}
where $\mathcal{F}'_\pm(k;x)\equiv\partial\mathcal{F}_\pm(k;x)/\partial x$.

By differentiating Eq.~(\ref{eqb-165}), we have
\begin{align}\label{eqb-180}
\phi'(k;x)&=\frac{1}{2ik}\left(\mathcal{F}'_-(k;x)e^{ikx}-\mathcal{F}'_+(k;x)e^{-ikx}\right)
\nonumber\\
&+\frac{ik}{2ik}\left(\mathcal{F}_-(k;x)e^{ikx}+\mathcal{F}_+(k;x)e^{-ikx}\right)
\\ 
\label{eqb-181}
&=\frac{ik}{2ik}\left(\mathcal{F}_-(k;x)e^{ikx}+\mathcal{F}_+(k;x)e^{-ikx}\right),
\end{align}
where we used the gauge-fixing condition~(\ref{eqb-170}) in moving from Eq.~(\ref{eqb-180}) to Eq.~(\ref{eqb-181}).
By differentiating Eq.~(\ref{eqb-181}), we have
\begin{align}\label{eqb-190}
\phi''(k;x)&=\frac{ik}{2ik}\left(\mathcal{F}'_-(k;x)e^{ikx}+\mathcal{F}'_+(k;x)e^{-ikx}\right)
\nonumber\\
&+\frac{-k^2}{2ik}\left(\mathcal{F}_-(k;x)e^{ikx}-\mathcal{F}_+(k;x)e^{-ikx}\right)
\\
\label{eqb-191}
&=\frac{1}{2}\left(\mathcal{F}'_-(k;x)e^{ikx}+\mathcal{F}'_+(k;x)e^{-ikx}\right)
-k^2\phi(k;x).
\end{align}
Comparing this with the Schr\"{o}dinger equation~(\ref{eq1-5}), we have
\begin{align}\label{eqb-200}
\frac{1}{2}\left(\mathcal{F}'_-(k;x)e^{ikx}+\mathcal{F}'_+(k;x)e^{-ikx}\right)
=\frac{V(x)}{2ik}\left(\mathcal{F}_-(k;x)e^{ikx}-\mathcal{F}_+(k;x)e^{-ikx}\right).
\end{align}

Equations~(\ref{eqb-170}) and~(\ref{eqb-200}) constitute a set of first-order differential equations for the functions $\mathcal{F}_\pm(k;x)$.
By solving them with respect to $\mathcal{F}'_\pm(k;x)$, we arrive at 
\begin{align}\label{eqb-210}
\mathcal{F}'_\pm(k;x)=e^{\pm ikx}\frac{V(x)}{2ik}\left(\mathcal{F}_-(k;x)e^{ikx}-\mathcal{F}_+(k;x)e^{-ikx}\right).
\end{align}
The boundary conditions~(\ref{eqa-30}) and~(\ref{eqa-31}) are now translated into the conditions
\begin{align}\label{eqb-220}
-\mathcal{F}_+(k;0)=\mathcal{F}_-(k;0)=ik &\quad\mbox{for even solutions,}
\\ \label{eqb-221}
\mathcal{F}_+(k;0)=\mathcal{F}_-(k;0)=1 & \quad\mbox{for odd solutions.}
\end{align}

We first solve the set of differential equations~(\ref{eqb-210}) from $x=0$ with the boundary conditions~(\ref{eqb-220}) and~(\ref{eqb-221}) up until a large enough $x=x_\mathrm{max}$, scanning the complex $k$ plane.
We then make a density plot of $|\mathcal{F}_+(k;x_\mathrm{max})|$ in the complex $k$ plane.
The density plot shows dips, which give rough estimates of the zeros of the function $\mathcal{F}_+(k;x_\mathrm{max})$.
Then we can use the standard Newton method to find each zero more precisely.
In other words, we numerically solve the equation
\begin{align}\label{eqb-230}
\mathcal{F}_+(k_\mathrm{res};x_\mathrm{max})=0
\end{align}
instead of Eq.~(\ref{eqb-168}).
This is how we obtained the results in Fig.~\ref{fig5}, where we used $x_\mathrm{res}=15/a_0$ after confirming that the function $\mathcal{F}_+(k_\mathrm{res};x_\mathrm{max})$ converges to a constant for $x\geq15/a_0$.

\end{document}